\documentclass[aps,prl,superscriptaddress,amsmath,amssymb,floatfix,reprint]{revtex4-2}
\pdfoutput=1

\usepackage{amssymb}
\usepackage{graphicx}
\usepackage{bm}
\usepackage[pdftex,breaklinks=true,bookmarksopen=true,bookmarksopenlevel=3,bookmarksnumbered=true,colorlinks=true,urlcolor= magenta,citecolor=blue,linkcolor=blue]{hyperref}
\usepackage{float}
\usepackage[utf8]{inputenc}
\usepackage[T1]{fontenc}
\usepackage{verbatim}
\usepackage{siunitx}
\usepackage[dvipsnames]{xcolor}
\usepackage{soul}
\usepackage{ulem}

\begin{document}

\title{Compton photons at the GeV scale from self-aligned collisions with a plasma mirror}

\author{Aimé Matheron}
\thanks{These authors contributed equally: Aimé Matheron, Jean-Raphaël Marquès, Vincent Lelasseux, Yinren Shou.}
\affiliation{Laboratoire d'Optique Appliquée, ENSTA Paris, CNRS, École Polytechnique, Institut Polytechnique de Paris, 91762 Palaiseau, France}

\author{Jean-Raphaël Marquès}
\thanks{These authors contributed equally: Aimé Matheron, Jean-Raphaël Marquès, Vincent Lelasseux, Yinren Shou.}
\affiliation{LULI, CNRS, École Polytechnique, CEA, Sorbonne Université, Institut Polytechnique de Paris, 91128 Palaiseau, France}

\author{Vincent Lelasseux}
\thanks{These authors contributed equally: Aimé Matheron, Jean-Raphaël Marquès, Vincent Lelasseux, Yinren Shou.}
\affiliation{Extreme Light Infrastructure - Nuclear Physics, Horia Hulubei National Institute for Physics and Nuclear Engineering - Str. Reactorului 30, Bucharest-Măgurele 077125, Romania}

\author{Yinren Shou}
\thanks{These authors contributed equally: Aimé Matheron, Jean-Raphaël Marquès, Vincent Lelasseux, Yinren Shou.}
\affiliation{Weizmann Institute of Science, 7610001 Rehovot, Israel}

\author{Igor A. Andriyash}
\affiliation{Laboratoire d'Optique Appliquée, ENSTA Paris, CNRS, École Polytechnique, Institut Polytechnique de Paris, 91762 Palaiseau, France}

\author{Vanessa~Ling Jen Phung}
\affiliation{Extreme Light Infrastructure - Nuclear Physics, Horia Hulubei National Institute for Physics and Nuclear Engineering - Str. Reactorului 30, Bucharest-Măgurele 077125, Romania}

\author{Yohann Ayoul}
\affiliation{LULI, CNRS, École Polytechnique, CEA, Sorbonne Université, Institut Polytechnique de Paris, 91128 Palaiseau, France}

\author{Audrey Beluze}
\affiliation{LULI, CNRS, École Polytechnique, CEA, Sorbonne Université, Institut Polytechnique de Paris, 91128 Palaiseau, France}

\author{Ioan Dăncuş}
\affiliation{Extreme Light Infrastructure - Nuclear Physics, Horia Hulubei National Institute for Physics and Nuclear Engineering - Str. Reactorului 30, Bucharest-Măgurele 077125, Romania}

\author{Fabien Dorchies}
\affiliation{Université de Bordeaux, CNRS, CEA, Centre Lasers Intenses et Applications, UMR 5107, 33400 Talence, France}

\author{Flanish~D'Souza}
\affiliation{Extreme Light Infrastructure - Nuclear Physics, Horia Hulubei National Institute for Physics and Nuclear Engineering - Str. Reactorului 30, Bucharest-Măgurele 077125, Romania}
\affiliation{Department of Physics, Lund University, P.O. Box 118, SE-22100, Lund, Sweden}

\author{Mathieu Dumergue}
\affiliation{LULI, CNRS, École Polytechnique, CEA, Sorbonne Université, Institut Polytechnique de Paris, 91128 Palaiseau, France}

\author{Mickaël Frotin}
\affiliation{LULI, CNRS, École Polytechnique, CEA, Sorbonne Université, Institut Polytechnique de Paris, 91128 Palaiseau, France}

\author{Julien Gautier}
\affiliation{Laboratoire d'Optique Appliquée, ENSTA Paris, CNRS, École Polytechnique, Institut Polytechnique de Paris, 91762 Palaiseau, France}

\author{Fabrice Gobert}
\affiliation{LULI, CNRS, École Polytechnique, CEA, Sorbonne Université, Institut Polytechnique de Paris, 91128 Palaiseau, France}

\author{Marius~Gugiu}
\affiliation{Extreme Light Infrastructure - Nuclear Physics, Horia Hulubei National Institute for Physics and Nuclear Engineering - Str. Reactorului 30, Bucharest-Măgurele 077125, Romania}

\author{Santhosh~Krishnamurthy}
\affiliation{Weizmann Institute of Science, 7610001 Rehovot, Israel}

\author{Ivan Kargapolov}
\affiliation{Weizmann Institute of Science, 7610001 Rehovot, Israel}

\author{Eyal Kroupp}
\affiliation{Weizmann Institute of Science, 7610001 Rehovot, Israel}

\author{Livia Lancia}
\affiliation{LULI, CNRS, École Polytechnique, CEA, Sorbonne Université, Institut Polytechnique de Paris, 91128 Palaiseau, France}

\author{Alexandru~Laz\u ar} 
\affiliation{Extreme Light Infrastructure - Nuclear Physics, Horia Hulubei National Institute for Physics and Nuclear Engineering - Str. Reactorului 30, Bucharest-Măgurele 077125, Romania}

\author{Adrien~Leblanc}
\affiliation{Laboratoire d'Optique Appliquée, ENSTA Paris, CNRS, École Polytechnique, Institut Polytechnique de Paris, 91762 Palaiseau, France}

\author{Mohamed Lo}
\affiliation{LULI, CNRS, École Polytechnique, CEA, Sorbonne Université, Institut Polytechnique de Paris, 91128 Palaiseau, France}

\author{Damien Mataja}
\affiliation{LULI, CNRS, École Polytechnique, CEA, Sorbonne Université, Institut Polytechnique de Paris, 91128 Palaiseau, France}

\author{François Mathieu}
\affiliation{LULI, CNRS, École Polytechnique, CEA, Sorbonne Université, Institut Polytechnique de Paris, 91128 Palaiseau, France}

\author{Dimitrios Papadopoulos}
\affiliation{LULI, CNRS, École Polytechnique, CEA, Sorbonne Université, Institut Polytechnique de Paris, 91128 Palaiseau, France}

\author{Pablo~San~Miguel~Claveria}
\affiliation{Instituto Superior Técnico, Universidade de Lisboa, 1049-001 Lisbon, Portugal}

\author{Kim Ta Phuoc}
\email{kim.ta-phuoc@cnrs.fr}
\affiliation{Université de Bordeaux, CNRS, CEA, Centre Lasers Intenses et Applications, UMR 5107, 33400 Talence, France}

\author{Anda-Maria Talposi}
\affiliation{Weizmann Institute of Science, 7610001 Rehovot, Israel}

\author{Sheroy Tata}
\affiliation{Weizmann Institute of Science, 7610001 Rehovot, Israel}

\author{Călin A. Ur}
\affiliation{Extreme Light Infrastructure - Nuclear Physics, Horia Hulubei National Institute for Physics and Nuclear Engineering - Str. Reactorului 30, Bucharest-Măgurele 077125, Romania}

\author{Daniel~Ursescu}
\affiliation{Extreme Light Infrastructure - Nuclear Physics, Horia Hulubei National Institute for Physics and Nuclear Engineering - Str. Reactorului 30, Bucharest-Măgurele 077125, Romania}

\author{Lidia V\u asescu} 
\affiliation{Extreme Light Infrastructure - Nuclear Physics, Horia Hulubei National Institute for Physics and Nuclear Engineering - Str. Reactorului 30, Bucharest-Măgurele 077125, Romania}

\author{Domenico Doria}
\affiliation{Extreme Light Infrastructure - Nuclear Physics, Horia Hulubei National Institute for Physics and Nuclear Engineering - Str. Reactorului 30, Bucharest-Măgurele 077125, Romania}

\author{Victor Malka}
\email{victor.malka@weizmann.ac.il}
\affiliation{Extreme Light Infrastructure - Nuclear Physics, Horia Hulubei National Institute for Physics and Nuclear Engineering - Str. Reactorului 30, Bucharest-Măgurele 077125, Romania}
\affiliation{Weizmann Institute of Science, 7610001 Rehovot, Israel}

\author{Petru Ghenuche}
\email{petru.ghenuche@eli-np.ro}
\affiliation{Extreme Light Infrastructure - Nuclear Physics, Horia Hulubei National Institute for Physics and Nuclear Engineering - Str. Reactorului 30, Bucharest-Măgurele 077125, Romania}

\author{Sebastien Corde}
\email{sebastien.corde@polytechnique.edu}
\affiliation{Laboratoire d'Optique Appliquée, ENSTA Paris, CNRS, École Polytechnique, Institut Polytechnique de Paris, 91762 Palaiseau, France}
\affiliation{SLAC National Accelerator Laboratory, Menlo Park, California 94025, USA}

\begin{abstract}

With today's multi-petawatt lasers, testing quantum electrodynamics (QED) in the strong field regime, where the electric field exceeds the Schwinger critical field in the rest frame of an electron, becomes within reach. Inverse Compton scattering of an intense laser pulse off a high-energy electron beam is the mainstream approach, resulting in the emission of high-energy photons that can decay into Breit-Wheeler electron-positron pairs. Here, we demonstrate experimentally that very high energy photons can be generated in a self-aligned single-laser Compton scattering setup, combining a laser-plasma accelerator and a plasma mirror. Reaching up to the GeV scale, photon emission via nonlinear Compton scattering exhibits a nonclassical scaling in the experiment that is consistent with electric fields reaching up to a fraction $\chi\simeq0.3$ of the Schwinger field in the electron rest frame. These foolproof collisions guaranteed by automatic laser-electron overlap provide a new approach for precise investigations of strong-field QED processes.

\end{abstract}

\maketitle


Multi-petawatt (PW) lasers are emerging worldwide \cite{Danson_HPLSE_2015, Danson2019, Yoon_Optica_2021, Radier_HPLSE_2022}, reviving interest for experimental investigations of strong-field quantum electrodynamics (QED) processes using all-optical approaches \cite{Nikishov_JETP_1964, Erber_RMP_1966, DiPiazza_RMP_2012}. These concepts usually involve electron beams from laser-plasma accelerators (LPA)
\cite{Tajima_PRL_1979,Faure_Nature_2004, Geddes_Nature_2004, Mangles_Nature_2004}, now capable of reaching electron energies up to 10 GeV \cite{Gonsalves2019, Aniculaesei2023, Picksley2024}), that collide with an intense counterpropagating laser pulse. They have enabled measurements of radiation reaction \cite{Cole_PRX_2018,Poder_PRX_2018} and nonlinear Compton scattering \cite{Mirzaie2024}, complementing the results obtained nearly three decades ago in the E-144 experiment with the 47 GeV electron beam from the SLAC linear accelerator~\cite{Bula1996,Burke1997,Bamber1999}, withhundred positrons observed and attributed to the multiphoton Breit-Wheeler reaction. In these experiments, the high-energy electrons experienced electric fields up to a fraction $\chi\simeq0.46$ of the Schwinger critical electric field $E_S=m_e^2c^3/e\hbar\simeq \SI{1.3e18}{V.m^{-1}}$, where $\chi=E^\star/E_S$ is the electron quantum parameter and $E^\star$ the electric field in the electron rest frame, $c$ the speed of light, $\hbar$ the reduced Plank constant, $e$ and $m_e$ the electron charge and mass.

\begin{figure*}[t!] 
  \centering
  \includegraphics[width=16cm]{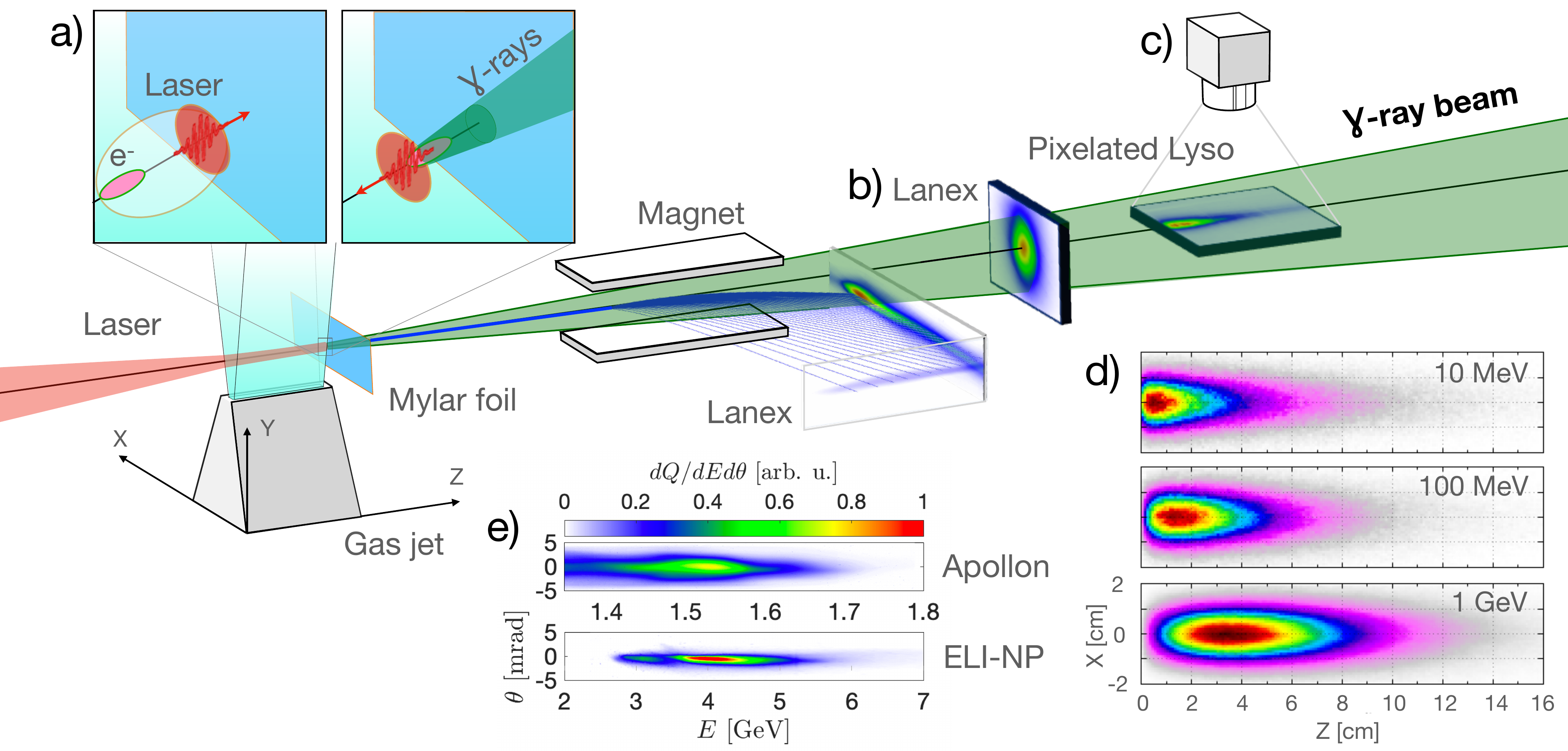} 
  \caption{\textbf{Principle of the plasma-mirror Compton scattering concept and experimental setup.} (a) The laser is focused onto a gas jet to drive a laser-plasma accelerator, and then reflected by a Mylar foil acting as a plasma mirror. The collision between LPA electrons and the reflected laser leads to the emission of high-energy Compton photons that are measured with a Lanex scintillator (b, transverse distribution) and a pixelated LYSO crystal array (c, energy deposition profile). (d) Simulated energy deposition profiles in the pixelated LYSO are shown for broadband photon spectra with critical energy $\hbar\omega_c=\SI{10}{MeV}$, $\hbar\omega_c=\SI{100}{MeV}$ and $\hbar\omega_c=\SI{1}{GeV}$. The LPA electron energy spectrum (e) is measured with a magnet and Lanex scintillators. }
  \label{FIGURE-Setup}
\end{figure*}

While these all-optical Compton scattering sources \cite{Schwoerer2006,Mori_APEX_2012,Chen_PhysRevLett_2013,Sarri_PhysRevLett_2014,Powers_NatPhoton_2014,Liu2014,Khrennikov2015}, as well as other on-going or planned accelerator-based Compton scattering experiments \cite{Yakimenko2019, Clarke2022, LUXE}, are requiring a sensitive overlap in time and space of the electron beam and the colliding laser pulse, the self-aligned Compton scattering approach has been realized by combining the LPA with a plasma mirror \cite{TaPhuoc2012,Tsai2015}. The plasma-mirror Compton scattering concept is depicted in Fig.~\ref{FIGURE-Setup}(a). A single intense laser pulse propagates into a quickly ionized gas jet target and drives a high-amplitude plasma wave, intense enough to trap and accelerate electrons to high energies over the length of the gas jet. A thin foil is placed at the output of the LPA. Ionized by the pedestal of the laser pulse, it turns into a plasma mirror. The laser pulse is backreflected towards the electrons and the collision results in the emission of high-energy Compton photons. In this scheme, the normalized vector potential of the reflected laser pulse, ${a_0=eE_l/m_ec\omega_l}$ ($E_l$ and $\omega_l$ being the laser peak electric field and central frequency), is typically larger than 1 and Compton scattering occurs in the nonlinear regime. The emitted radiation has a broadband spectrum, that can be characterized by a critical photon energy $\hbar\omega_c$ (see Methods) scaling as $\gamma_e^2a_0$ in the classical regime \cite{Corde2013} (with $\gamma_e$ the electron Lorentz factor), while quantum effects and the recoil associated with photon emission slow down this scaling and result in lower photon energies \cite{Ritus1985}.

In this paper, we demonstrate experimentally that the photon energy produced using this plasma-mirror Compton scattering concept can be scaled up from the sub-MeV range of the seminal result in Ref. \cite{TaPhuoc2012} to the GeV range, making it very relevant for the study of strong-field QED. Using Apollon 1~PW and ELI-NP 10~PW laser facilities, we respectively measured critical photon energies of ${\hbar\omega_c = 0.14\pm0.03\:\si{GeV}}$ and ${\hbar\omega_c = 0.55\pm0.1\:\si{GeV}}$. Remarkably, this successful scaling over 3 orders of magnitude (MeV to GeV) slows down as we approach the GeV range and enter the moderately quantum regime, as evidenced by the deviation of the measured critical photon energies $\hbar\omega_c$ at Apollon and ELI-NP with respect to the classical $\gamma_e^2a_0$ scaling. The results show that strong-field QED processes such as nonlinear Compton scattering can be probed in a single-laser setup with self-aligned collisions and \SI{100}{\%} collision success rate, free from the misalignment errors and fluctuations of usual multibeam approaches \cite{Mirzaie2024}.

The experiments were carried out at the Apollon \cite{APOLLON_Papadopoulos_HPLSE_2016} and ELI-NP \cite{Lureau2020} laser facilities, that respectively deliver laser pulses with an on-target energies of 15~J and 180~J and a pulse duration of 25~fs (full-width-at-half-maximum, FWHM). The experimental setup is presented in Fig.~\ref{FIGURE-Setup} and its components are described in detail in the Methods section. The laser pulse is focused with a spherical mirror with a f-number of 40 (Apollon) and 60 (ELI-NP) onto a gas jet (2-cm / 6-cm long, respectively), operating with nitrogen-doped helium to trigger ionization injection in the LPA \cite{McGuffey2010,Pak2010}. The laser strength parameters in vacuum are respectively ${a_0 \sim 3}$ and ${a_0 \sim 6}$ at Apollon and ELI-NP. The plasma mirror consists in a $\SI{250}{\mu m}$-thick Mylar foil that can be moved along the propagation axis and its position is denoted $Z_f$. We use a fresh surface for each shot. Electrons from the LPA are deviated by a meter-long magnet and detected with scintillating screens, so as to measure the electron energy spectrum in the GeV range. Typical electron energy spectra obtained  are shown in Fig.~\ref{FIGURE-Setup}(e) for Apollon and for ELI-NP. The Apollon electron beam has a maximum energy of about \SI{1.6}{GeV}, a charge of \SI{90}{pC} above \SI{1}{GeV} and a divergence along the vertical axis ($Y$ axis) of \SI{3.7}{mrad} (FWHM). The ELI-NP electron beam has a maximum energy of about \SI{5}{GeV}, a charge of \SI{1}{nC} above \SI{2}{GeV} and a divergence along the horizontal axis ($X$ axis) of \SI{1.6}{mrad} (FWHM). Additional electron spectra are shown in Supplementary Figs. S1 and S2 for Apollon and ELI-NP LPAs respectively. Compton gamma rays are detected after the electron spectrometer using two diagnostics located 2.5~meters (Apollon) and 7.4~meters (ELI-NP) from the source. A 2.5-mm-thick Cu converter foil immediately followed by a scintillating Lanex screen is used to measure the angular distribution and total flux of the Compton gamma-ray beam [Fig.~\ref{FIGURE-Setup}(b)]. The Cu converter enhances the sensitivity to high-energy photons (above the MeV), while blocking lower-energy photons, in particular betatron radiation from the LPA \cite{Rousse2004}. A pixelated LYSO:Ce crystal array oriented in the $Z-X$ plane is used to capture the spectral properties of the Compton gamma-ray beam [Fig.~\ref{FIGURE-Setup}(c)]. The simulated energy deposition profiles, depicted in Fig. \ref{FIGURE-Setup}(d), show the sensitivity of this gamma-ray detector to different critical photon energies~$\hbar\omega_c$.


\begin{figure}[t!]
 \centering
  \includegraphics[width=\columnwidth]{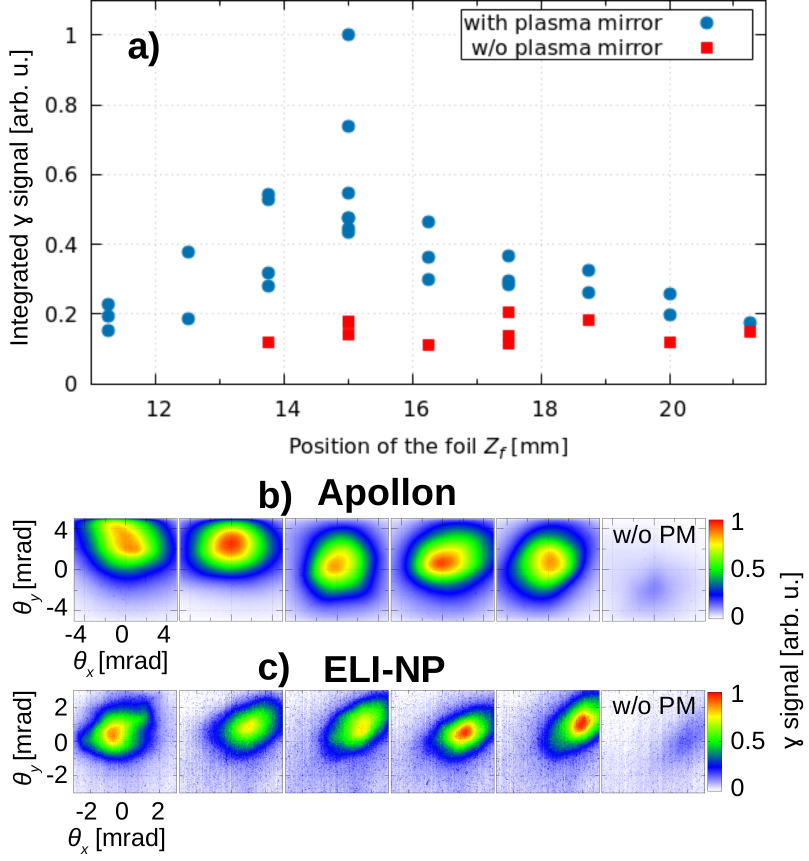}
 \caption{\textbf{Evidence of Compton gamma rays and collision success rate.} (a) Gamma-ray signal in the Apollon experiment, obtained by integrating the signal on the gamma-ray Lanex [(b) in Fig.~\ref{FIGURE-Setup}], and measured as a function of the foil position $Z_f$ using fresh foil (with plasma mirror, Compton shots taken after rotating the foil) and by shooting in a hole (without plasma mirror, reference shot taken just after a Compton shot without rotating the foil). (b)-(c) Angular distribution of the Compton gamma-ray beam for successive shots around the optimum for the Apollon (b) and ELI-NP (c) experiments, as well as bremsstrahlung background shot (shot in a hole, marked as ``w/o PM'').}
 \label{FIGURE-ZSCAN}
\end{figure}

To demonstrate the generation of Compton gamma rays with this plasma-mirror scheme, the experimental evidence is based on two measurements: the variation of the gamma-ray signal with the plasma mirror position (as the Compton radiation depends on the electron energy and the intensity of the backreflected laser) and the comparison to the signal observed without plasma mirror, as shown in Fig. \ref{FIGURE-ZSCAN}. We start with the Apollon experiment. When the foil is located at a distance between ${Z_f\sim\SI{10}{mm}}$ and ${Z_f=\SI{15}{mm}}$ from the beginning of the gas jet (defined as $Z_f=0$), the gamma-ray signal increases with $Z_f$ as expected from the increase of the LPA acceleration length that leads to higher electron energies and charges. Beyond ${Z_f=\SI{15}{mm}}$, electrons no longer evolve substantially and the gamma-ray signal drops, suggesting that the laser pulse starts to diffract and does not efficiently drive the LPA after 15 mm. This specific dependence of the gamma-ray signal with the foil position [Fig.~\ref{FIGURE-ZSCAN}(a)], characteristic of the plasma-mirror Compton scheme~\cite{TaPhuoc2012}, is the first evidence that Compton gamma rays have been generated and observed in our experiment.

Second, one can compare Compton shots with plasma mirror to reference shots without plasma mirror. The reference shots are taken with the foil inserted at a given $Z_f$ position, but with a hole along the laser axis generated by the previous Compton shot. This allows to have a similar LPA gas profile for reference shots. Without plasma mirror, one expects a $Z_f$-independent background signal from bremsstrahlung radiation generated by the electron beam passing through foils before being deflected by the magnet, which is indeed observed in the experiment. Figures~\ref{FIGURE-ZSCAN}(a)-(b) show that Compton shots around the optimum ${Z_f=\SI{15}{mm}}$ position are much brighter than the bremsstrahlung background in the Apollon experiment, and a similar result is observed for the ELI-NP experiment in Fig.~\ref{FIGURE-ZSCAN}(c). The bremsstrahlung background was found to be sufficiently small to not impact the spectral analysis discussed below. With the higher laser energy and the longer cm-scale Rayleigth length in the ELI-NP experiment, Compton gamma-ray generation was better optimized by having the plasma mirror after the gas jet exit, at a distance of 2 to 3 cm. As anticipated from this self-aligned concept, all shots with a plasma mirror resulted in a successful collision, as depicted in Figs.~\ref{FIGURE-ZSCAN}(b)-(c) showing five successive shots with \SI{100}{\%} success rate for each experiment.


\begin{figure*}[t!]
	\centering
	\includegraphics[width=0.9\textwidth]{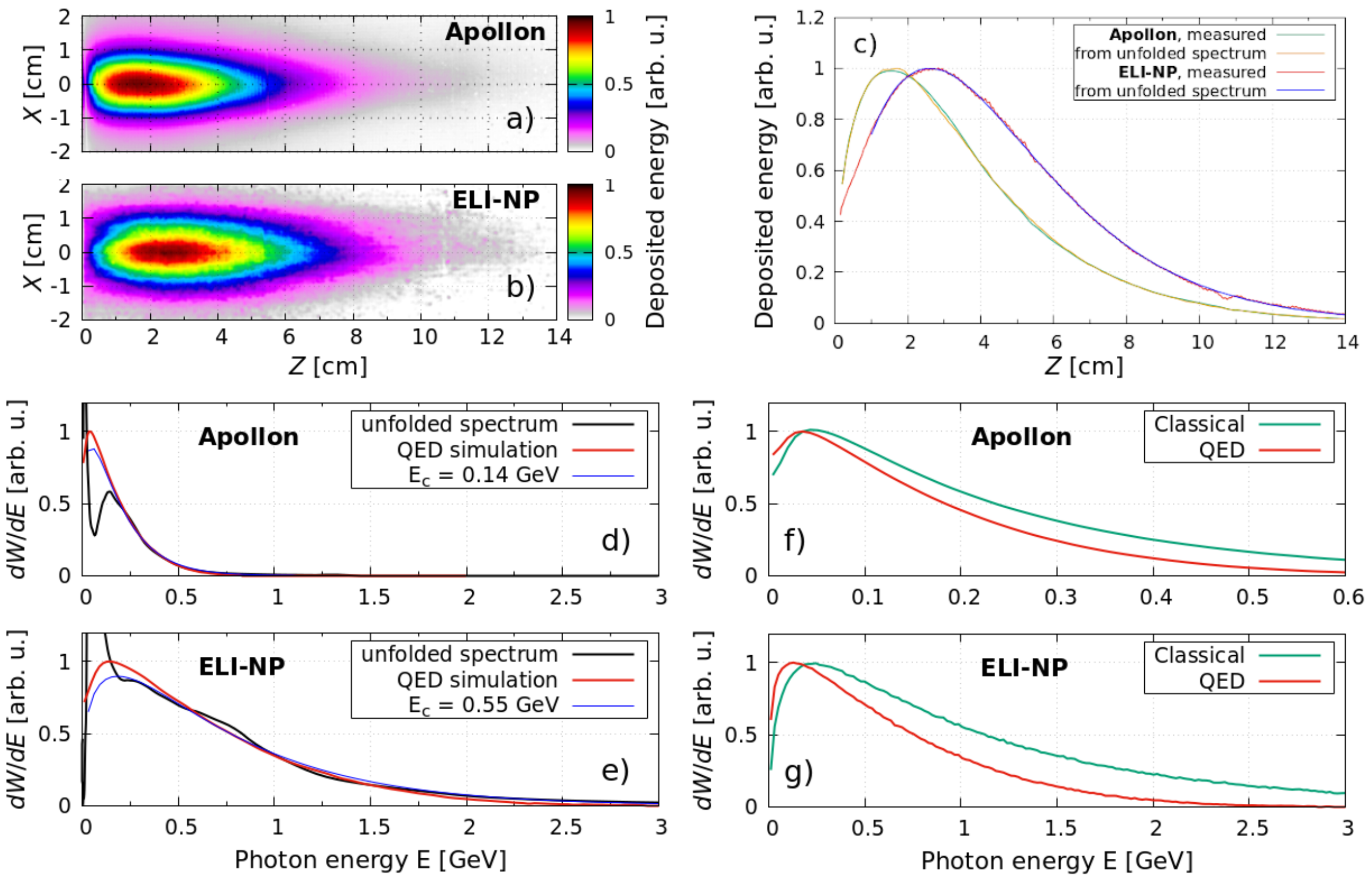}
	\caption{\textbf{Energy deposition in LYSO and Compton photon spectrum.} Experimental 2D profiles of the energy deposition in LYSO by a Compton gamma-ray beam at Apollon (a) and ELI-NP (b). (c) 1D depth profiles obtained by integration of (a) and (b) along $X$, as well as depth profiles from spectral reconstructions. (d)-(e) Experimental photon spectra from iterative spectral reconstructions, from synchrotron fits with critical energies $\hbar\omega_c=\SI{0.14}{GeV}$ (Apollon) and $\hbar\omega_c=\SI{0.55}{GeV}$ (ELI-NP) and from QED simulations. Simulated photon spectra in classical (including continuous radiation reaction) and QED regimes, for Apollon (f) and ELI-NP (g) conditions.}
	\label{FIGURE-LYSO}
\end{figure*}

Figures~\ref{FIGURE-LYSO}(a)-(b) present experimental images of the pixelated LYSO detector at Apollon and ELI-NP. The maximum of the 1D energy deposition profile [Fig.~\ref{FIGURE-LYSO}(c)] clearly shifts from ${Z=\SI{1.5}{cm}}$ (Apollon) to ${Z=\SI{2.7}{cm}}$ (ELI-NP), which reflects the difference in the photon energy spectra. This detector indeed encodes spectral information of the gamma photons via energy-dependent deposition [see Fig.~\ref{FIGURE-Setup}(d) and Methods] that we have characterized with Monte-Carlo simulations using the FLUKA code \cite{FLUKA_1,FLUKA_2,FLAIR}. With the measured energy deposition profile and known detector response, gamma-ray photon spectra can be unfolded and reconstructed by an iterative method using the Maximum Likelihood Expectation Maximization algorithm (see Methods). The experimental energy deposition profiles and the unfolded photon spectra obtained from the reconstruction method are shown in Figs. \ref{FIGURE-LYSO}(c)-(e). Figures \ref{FIGURE-LYSO}(d)-(e) also show the results of QED simulations of the laser-electron collision, as well as synchrotron fits to the experimental unfolded photon spectra, yielding $\hbar\omega_c=0.14\pm0.03\:\si{GeV}$ (Apollon) and $\hbar\omega_c=0.55\pm0.1\:\si{GeV}$ (ELI-NP). The simulations use the strong-field QED code Ptarmigan \cite{ptarmigan} and determine the incoming electron energy spectrum to match the outcoming spectrum to the experimental one. The QED simulation includes the effects of discrete photon emission and of the electron energy loss and recoil for multiple photon emissions from the same electron. A reasonable agreement is found between the experiment and the QED simulation using a laser strength parameter $a_0=8$ for the Apollon Compton shot in Fig. \ref{FIGURE-LYSO}(a) and $a_0=5$ for the ELI-NP Compton shot in Fig. \ref{FIGURE-LYSO}(b). The larger $a_0$ in the Apollon case is consistent with the sensitive foil positioning inside the gas jet, and with the laser self-focusing in the plasma resulting in smaller spot size and Rayleigth length. In the ELI-NP case, the foil positioning after the gas jet exit and the long cm-scale Rayleigh length lead to a more moderate value for $a_0$. 

The increase in the critical photon energy, from $\hbar\omega_c=\SI{0.14}{GeV}$ at Apollon to $\hbar\omega_c=\SI{0.55}{GeV}$ at ELI-NP, is expected from the increase in electron energy from a maximum of about \SI{1.6}{GeV} to \SI{5}{GeV}. However, the observed increase in $\hbar\omega_c$ by a factor of nearly 4 exhibits a substantial deviation from the classical $\gamma_e^2a_0$ scaling predicting a $\times6$ increase. This is due to quantum effects and to the recoil associated with photon emission that become important, especially for the ELI-NP experiment. Indeed, our experiments are entering the moderately quantum regime with $\chi\simeq0.15$ (Apollon, corresponding to 1.6 GeV electron energy and $a_0=8$) and $\chi\simeq0.3$ (ELI-NP, corresponding to 5 GeV electron energy and $a_0=5$) where the quantum nature of the interaction needs to be taken into account \cite{Ritus1985} and clearly deviates from classical theory, as shown in Figs. \ref{FIGURE-LYSO}(f)-(g). Simulated photon spectra are compared for classical and QED regimes (see Methods), showing that the classical-quantum deviation is more prominent for ELI-NP [Figs. \ref{FIGURE-LYSO}(g)] than for Apollon [Figs. \ref{FIGURE-LYSO}(f)], slowing down the growth in $\hbar\omega_c$. 

Finally, using the measured electron charge and the results from the ELI-NP QED simulation, we estimate the number of Compton gamma photons produced at energies beyond \SI{1}{GeV} to be of the order of $10^8$ in the ELI-NP experiment. This is similar to the estimated number of photons at energies beyond \SI{10}{MeV} in multibeam Compton experiment~\cite{Mirzaie2024}, while benefiting from automatic alignment and foolproof collisions with a  critical energy $\hbar\omega_c$ about four times larger.

In conclusion, Compton gamma-ray beams have been produced with photon energies exceeding 1 GeV in a simplified single-laser Compton scattering geometry and with \SI{100}{\%} collision success rate, that compares favorably with respect to usual multibeam approaches. For this scheme, the photon energy of the Compton radiation has been increased by about three orders of magnitude as compared to the initial seminal result in Ref. \cite{TaPhuoc2012}, entering the moderately quantum regime with $\chi$ up to $\simeq0.3$ and exhibiting a nonclassical scaling for the photon energy. The results open the way to strong-field QED investigations with Compton collisions free from shot-to-shot fluctuations associated to laser-electron overlap, benefiting from automatic alignment and synchronization provided by the plasma mirror. They also pave the way to pure light-by-light scattering experiments by deflecting electrons away and colliding GeV photons with an additional multi-PW laser.

\bigskip
\noindent
{\textbf{Methods}}

{\small

\noindent
\textbf{Apollon laser system.} 
For the experiment carried out at the Apollon facility in October 2023, the Ti:sapphire laser system delivered pulses with 15 Joules on target, \SI{25}{fs} (FWHM) duration, at a central wavelength of \SI{810}{nm}, with a linear polarization (along the horizontal $X$ axis) and with a repetition rate of one shot per minute. The laser pulse with a diameter of \SI{14}{cm} is focused with a spherical mirror with a focal length of \SI{6}{m} onto the gas jet, with a vacuum spot size of \SI{40}{\mu m} (FWHM). The laser strength parameter in vacuum is estimated to be $a_0 \sim 3$.

\noindent
\textbf{ELI-NP laser system.} 
For the experiment carried out at the ELI-NP facility in October 2024, the Ti:sapphire laser system delivered pulses with 180 Joules on target, \SI{25}{fs} (FWHM) duration, at a central wavelength of \SI{810}{nm}, with a linear polarization (along the horizontal $X$ axis) and with a repetition rate of one shot per minute. The laser pulse with a diameter of \SI{50}{cm} is focused with a spherical mirror with a focal length of \SI{30.5}{m} onto the gas jet, with a vacuum spot size of \SI{57}{\mu m} (FWHM). The laser strength parameter in vacuum is estimated to be $a_0 \sim 6$.

\noindent
\textbf{Targets.} 
The laser-plasma accelerator consists of a gas jet operated with a solenoid valve at backing pressures of \SI{15}{bar} (Apollon) and \SI{40}{bar} (ELI-NP). The gas jet nozzle has a \SI{1}{mm}-wide slit-shaped exit, that is \SI{2}{cm} long at Apollon and \SI{6}{cm} long for ELI-NP along the laser $Z$ axis. The laser axis is 2.5 mm (Apollon) and 6 mm (ELI-NP) above the top of the gas jet nozzle. Electron injection in the LPA is obtained by using ionization injection \cite{McGuffey2010,Pak2010} with an helium gas doped with nitrogen (\SI{1}{\%} and \SI{2}{\%} doping, respectively for Apollon and ELI-NP). The plasma density is estimated to be $\approx 10^{18}\:\si{cm^{-3}}$ in the Apollon experiment and $\approx 7\times10^{17}\:\si{cm^{-3}}$ in the ELI-NP experiment. A \SI{250}{\mu m}-thick Mylar foil is used for the plasma mirror. The foil is mounted on a rotating wheel so that a fresh surface can be used to trigger a Compton collision for the next shot by rotating the wheel. The foil can be moved along the $Z$ axis, above the gas jet as well as after its exit. After a Compton shot where the high-energy laser hits the foil, the foil has a hole with a typical diameter ranging from one millimeter to a few millimeters. Sending the next shot without rotating the wheel results in a shot in the hole, a reference shot where Compton gamma rays are not present. Reference shots (in the hole) have a similar LPA gas profile to the Compton shots because the foil is inserted in both cases.

\noindent
\textbf{Electron and gamma-ray diagnostics.} 
The electron spectrometer consists of a permanent dipole magnet, with \SI{1.4}{T} over \SI{1.2}{m} (Apollon) and \SI{0.95}{T} over \SI{0.8}{m} (ELI-NP), that deflects electron depending of their energy towards scintillation screens imaged onto CDD and CMOS cameras. The energy calibration is obtained by calculating electron trajectories through the measured magnetic field distribution in the dipole and by using the measured position on the screen of the gamma-ray centroid, corresponding to the non-deviated (infinite electron energy) axis. Compton gamma rays are characterized, spatially and spectrally, using two dedicated detectors. The first gamma-ray detector [Fig. \ref{FIGURE-Setup}(b)] consists of a Lanex screen oriented at an angle of 45 degrees with respect to the gamma axis and imaged onto a 16-bit visible CCD camera. A \SI{2.5}{mm}-thick copper converter is placed immediately before the Lanex screen to improve its sensitivity to Compton gamma rays and block lower-energy photons, in particular betatron radiation from the LPA. The second gamma-ray detector [Fig. \ref{FIGURE-Setup}(c)] consists of a pixelated LYSO:Ce (cerium-doped lutetium-yttrium oxyorthosilicate) crystal array oriented along the horizontal $Z-X$ plane, with dimensions $\SI{10.8}{cm}\: (X)\times\SI{4}{mm}\: (Y)\times\SI{16.2}{cm}\:(Z)$, and imaged from the top by a scientific 16-bit CDD camera. The LYSO pixels have dimensions $\SI{1}{mm}\: (X)\times\SI{4}{mm}\: (Y)\times\SI{1}{mm}\:(Z)$ and are separated by \SI{0.08}{mm}-thick reflective films. The generation of background bremsstrahlung radiation by the LPA electron beam traversing foils is minimized as much as possible. At Apollon, the magnet of the electron spectrometer is isolated from the interaction chamber by a 0.5-mm-thick aluminum foil. At ELI-NP, a \SI{50}{\mu m} Al foil is used to block the laser followed by \SI{100}{\mu m} C + \SI{100}{\mu m} Kapton to separate the vacuum chamber from the electron and gamma detectors located outside the interaction chamber.

\noindent
\textbf{Spectral analysis.} 
The energy deposition along the $Z$ axis in the LYSO detector encodes spectral information by the photon-energy-dependent depth deposition profile. Using the 3D Monte-Carlo code FLUKA \cite{FLUKA_1, FLUKA_2, FLAIR} and the experimentally-measured gamma-ray divergence, the depth deposition response of the LYSO detector has been simulated for mono-energetic photons with energies ranging from 1 MeV to 5 GeV. This detector response is then used in the reconstruction of the photon spectrum using an iterative method based on a Maximum Likelihood Expectation Maximization algorithm, commonly used in tomography \cite{Shepp,Lange,Miller}. Without any assumption on the final spectral shape, a least square regression approach starts from a flat spectrum and iteratively unfolds the photon spectrum, minimizing the residual sum of squares (RSS) in the 1D depth deposition profiles of Fig.~\ref{FIGURE-LYSO}(c). The RSS is defined as
\begin{equation}
\nonumber
    RSS = \sum_{i=1}^n \left( D_i^c-D_i^m \right)^2,
\end{equation}
with $D_i^c$ and $D_i^m$ the calculated (from the reconstruction method) and measured value of the 1D depth deposition profile at pixel $i$ along the $Z$ axis. While the general spectral shape at high energies is robust, the reconstruction at small photon energies ($\lesssim\SI{100}{MeV}$) is sensitive to small imperfections or errors in the deposited energy profile and can result in unphysical oscillations at these low photon energies. To retrieve a critical photon energy $\hbar\omega_c$, the unfolded photon spectra are fitted with a synchrotron spectrum of the form $dW/d(\hbar\omega)\propto S(\omega/\omega_c)$, where $dW=\hbar\omega\: dN$ is the radiated energy and $dN$ the photon number in the energy band $d(\hbar\omega)$, $S(x)=x\int_x^{\infty}K_{5/3}(y)dy$ is the synchrotron function and $K_{5/3}$ is the modified Bessel function of the second kind.

\noindent
\textbf{Numerical modeling.}
In order to reproduce the process of nonlinear inverse Compton scattering, we have used the strong-field QED code Ptarmigan \cite{ptarmigan}. Ptarmigan is equipped with modules that calculate the radiation in the QED regime with the local constant-field approximation (LCFA) or with the locally-monochromatic approximation (LMA), and accounts for the discrete recoils associated to the emission of Compton photons. In Apollon and ELI-NP conditions, the LCFA and LMA simulations are in excellent agreement except at low photon energies where they slightly deviate (below \SI{50}{MeV} for Apollon and below \SI{300}{MeV} for ELI-NP). The QED simulations shown in Figs.~\ref{FIGURE-LYSO}(d)-(g) use LCFA. Ptarmigan can also calculate classical synchrotron radiation including continuous radiation reaction, and these classical results are shown in Figs.~\ref{FIGURE-LYSO}(f)-(g). The simulations of nonlinear Compton scattering are performed assuming a counterpropagating (reflected) laser pulse colliding with the electron beam in free space. The laser has Gaussian spatial and temporal profiles, a central wavelength of \SI{810}{nm}, a waist of \SI{50}{\mu m} (radius at $1/e^2$ in intensity), a pulse duration of \SI{25}{fs} (FWHM), and the laser strength parameter $a_0$ is varied to obtain a reasonable agreement between the QED simulations and the experimental measurements. The electron bunch being much smaller than the laser, its longitudinal and transverse profiles have no effect on the radiation properties, and we have considered a Gaussian bunch with \SI{1}{\mu m} size (root-mean-square, RMS) in all directions. The measured electron energy spectrum is used in the simulations by determining an input spectrum in the form of a Gaussian beams combination, such that the output spectrum of the QED simulation matches the experimental electron spectrum. The QED and classical simulations in Figs.~\ref{FIGURE-LYSO}(f)-(g) use the same input parameters, corresponding to Apollon [Fig.~\ref{FIGURE-LYSO}(f)] and ELI-NP [Fig.~\ref{FIGURE-LYSO}(g)] conditions, and only differ by the physics (QED with LCFA or classical including continuous radiation reaction) being modeled. The ELI-NP QED simulation uses the experimentally measured charge to obtain an order-of-magnitude estimate of the number of Compton photons generated above \SI{1}{GeV} in the ELI-NP experiment.
}

\bigskip
\noindent
\textbf{Data availability}

{\small
\noindent
The data that support the findings of this study are available from the corresponding authors upon reasonable request.
}

\bigskip
\noindent
\textbf{Code availability}

{\small
\noindent
FLUKA and Ptarmigan codes have been used to generate results presented in Figs. \ref{FIGURE-Setup}(e) and \ref{FIGURE-LYSO}. The FLUKA code is described in Refs. \cite{FLUKA_1,FLUKA_2,FLAIR} and is available open source at \href{https://fluka.cern}{https://fluka.cern}. The Ptarmigan code is described in Ref. \cite{ptarmigan} and is available open source at \href{https://github.com/tgblackburn/ptarmigan}{https://github.com/tgblackburn/ptarmigan}.
}

\noindent

%

\bigskip
\noindent
{\textbf{Acknowledgments}} 

\noindent
The authors acknowledge the national research infrastructure Apollon and LULI for their technical assistance in the Apollon experiment. We acknowledge ELI-NP technical staff for laser operation and technical assistance for the ELI-NP experiment. The work at LOA and LULI was supported by the ANR (g4QED project, Grant No. ANR-23-CE30-0011). AM was supported by the ANR under the program ``Investissements d'Avenir'' (Grant No. ANR-18-EURE-0014). The Extreme Light Infrastructure - Nuclear Physics (ELI-NP) team acknowledges the support of the Romanian Government and the European Union through the European Regional Development Fund - the Competitiveness Operational Program (1/07.07.2016, COP, ID 1334), the Romanian Ministry of Research and Innovation under research contract PN23210105 (Phase 2, the Program Nucleu) and PN23210106, IOSIN funds for research infrastructures of national interest and ELI-RO project LGS (Contract ELI-RO/RDI/2024\_021). FD’S acknowledges the support of the European Unions’s Horizon Europe research and innovation program under grant agreement no. 101073480. The Weizmann team acknowledges the support of the ISF, Minerva, PHC/IMOS. KT acknowledges the support of PHC/Maimonide program no. 50048WB. The authors are grateful to Cédric Thaury for fruitful discussions and help during the Apollon experiment.

\bigskip
\noindent
{\textbf{Author contributions}} 

\noindent
The experiment at Apollon and its data analysis were carried out by AM, JRM, LL, JG, IAA, FDo, PS, ALe, KT, SC with the supervision of KT and support of MF, ML, MD, FM. Laser shots at Apollon were provided by YA, AB, FG, DM, DP. The experiment at ELI-NP and its data analysis were carried out by PG, VL, YS, VP, SK, AT, IK, MG, FD’S, ST, EK, AM, JRM, LL, IAA, KT, SC, with the supervision of PG, DD, KT, SC, VM and support of CU. Laser shots at ELI-NP were provided by DU, ID, ALa, LV. Simulations and analytical calculations were carried out by IAA, AM. KT, SC, PG, VM proposed and managed the experiments. AM, JRM, IAA, KT, PG, VM, SC wrote the manuscript with inputs from all authors.

\bigskip
\noindent
{\textbf{Competing interests}} 

\noindent
The authors declare no competing interests.

\end{document}